# Research on Feature Extraction Data Processing System For MRI of Brain Diseases Based on Computer Deep Learning

Lingxi Xiao[1], Jinxin Hu[2], Yutian Yang[3], Yinqiu Feng[4], Zichao Li[5], Zexi Chen[6]

[1]Georgia Institute of Technology,USA,Lingxi.xiao@gatech.edu

[2]Arizona State University,USA,jinxinhu@asu.edu

[3]University of California, Davis,USA,yytyang@ucdavis.edu

[4]Columbia University,USA,yf2579@columbia.edu

[5]Canoakbit Alliance Inc,Canada,zichaoli@canoakbit.com

[6]North Carolina State University,USA,chenzx.hit09@gmail.com

*Abstract*—Most of the existing wavelet image processing techniques are carried out in the form of single-scale reconstruction and multiple iterations. However, processing high-quality fMRI data presents problems such as mixed noise and excessive computation time. This project proposes the use of matrix operations by combining mixed noise elimination methods with wavelet analysis to replace traditional iterative algorithms. Functional magnetic resonance imaging (fMRI) of the auditory cortex of a single subject is analyzed and compared to the wavelet domain signal processing technology based on repeated times and the world's most influential SPM8. Experiments show that this algorithm is the fastest in computing time, and its detection effect is comparable to the traditional iterative algorithm. However, this has a higher practical value for the processing of FMRI data. In addition, the wavelet analysis method proposed signal processing to speed up the calculation rate.

## I. Introduction

Functional magnetic resonance imaging (fMRI) is assuming a growing significance in the realm of non-invasive studies, primarily due to its exceptional temporal and spatial resolution, as well as its capacity to acquire anatomical and functional images concurrently. However, in fMRI, the brain response to experimental stimuli is weak, mixed with random noise and low-frequency drift. Therefore, FMRI technology needs to eliminate external influences while trying to avoid damaging the response of brain regions [1]. At present, the research of functional magnetic resonance imaging (fMRI) mainly includes pattern-based and data-based studies, among which the statistical parameter map is the most authoritative [2]. The high-pass filter was used to eliminate the shift of the low-frequency band, and the measured excitation function and the convolution of the hemodynamic equation were used to simulate and test the brain region response induced by stimulation [3]. A general linear mathematical model was established to model and statistically analyze the timing of each body element.

Wavelet transform can decompose time-frequency features into various wavelets, so it has an important application in biomedical information [4]. The multi-scale decomposition technology of wavelet signal is to reconstruct it in a certain wavelet domain by using the characteristics of this multi-scale decomposition, so as to process the signal in the signal [5]. Functional magnetic resonance imaging (fMRI) is studied by using wavelet multi-scale feature extraction. However, the signal processing algorithm in the wavelet image makes it easy to produce mixed interference in the process of single-scale reconstruction of large data such as functional magnetic resonance imaging, and its wide application [6]. The project intends to combine the hybrid denoising technology and wavelet transform, and build a new signal processing method based on wavelet domain through the matrix calculation of wavelet transform, so as to realize the accurate and rapid analysis of large capacity FMRI big data.

## II. MULTI-SCALE FEATURE EXTRACTION METHOD OF DISCRETE WAVELET TRANSFORM

The wavelet multi-scale feature extraction algorithm based on wavelet analysis takes the wavelet scale corresponding to the brain region response signal U as the characteristic wavelet scale and obtains the corresponding coefficients on the wavelet scale through the wavelet decomposition of U[7]. The multi-scale feature extraction algorithm based on wavelet analysis is mainly divided into the following three parts:

Step 1: Decompose the time series data $v$ of a voxel into a set of wavelet coefficients $z_v = [l_I, r_I, r_{I-1}, \cdots, r_1]$ distributed in $I+1$ wavelet scales based on the discrete wavelet transform Mallat algorithm

$$\begin{cases} l_0 = v \\ l_j = R_0 A_o l_{j-1} \\ r_j = R_0 B_i l_{j-1} \end{cases} \quad (1)$$

$A_o$ and $B_i$ are wavelet low-pass and high-pass decomposition filters. $R_0$ is a binary down sampling operator, which is used to collect even order terms of filter output. Decompose the series $j = 1, 2, \cdots, I$. $l_j$ And $r_j$ are called the $j$ low-frequency scale and high-frequency scale, respectively [8]. The formula (1) of order $I$ is expressed by the operator $E$, that is, $z_v = Ev$.

Step 2: Through the wavelet transform, judge the size of the wavelet domain where the brain region response signal contained in the activated body element is located [9]. The algorithm first decomposes U, then reconstructs each reconstruction on a single scale, and then compares the data from each reconstruction with the spectrum of U. Through spectrum analysis, it is determined that the characteristic wavelet scales of the hearing test are $r_1$ and $r_3$. Figure 1 shows the signal spectrum generated by the reconstruction of these two scales [10]. As can be seen from Figure 1, the multi-scale feature extraction method based on wavelet will produce mixed effects in the process of single-scale reconstruction, which is manifested as that the spectrum after single-scale reconstruction contains undetected frequency bands, which has a great impact on its wavelet scale identification [11]. In addition, the multi-scale feature extraction method based on wavelet achieves the whole information processing through iteration and repeated cycles, which is time-consuming and difficult to meet the needs of large-scale FMRI.

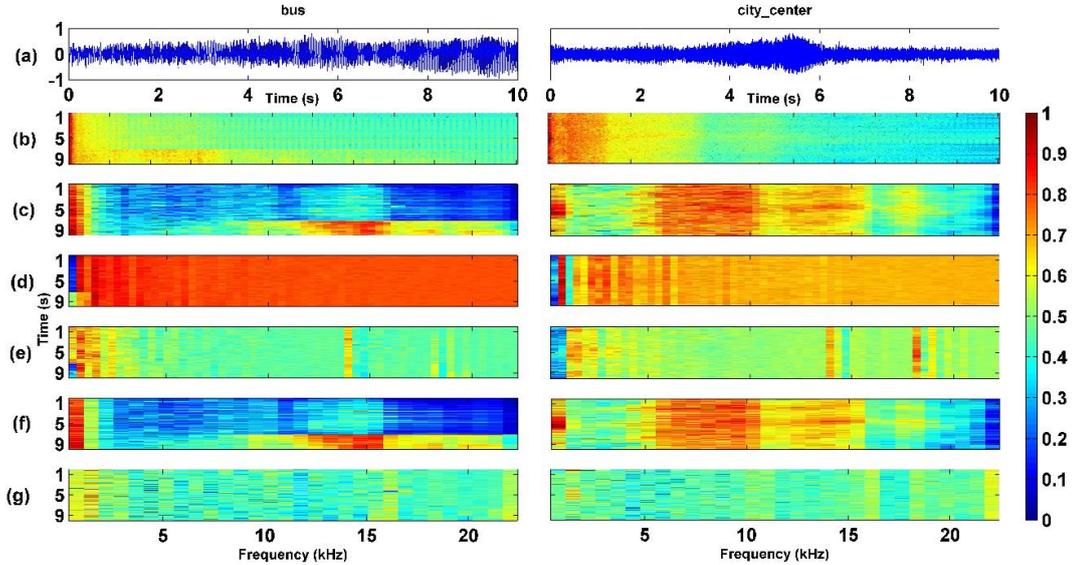

Fig. 1. Multi-scale feature extraction method of SPM8 and discrete wavelet transform analyzes the spectrum involved in auditory block fMRI data

Step 3: Keep the coefficients of the feature wavelet scale in the coefficient set $z_v$ unchanged, and set the coefficients of the other scales to zero.

Step 4: The modified set $z_v$ is reconstructed by discrete wavelet, and the voxel time series signal $v_E$ with the interference component removed is obtained

$$\begin{cases} l_{j-1} = a_o W l_j + b_i W r_j \\ v_E = l_0 \end{cases} \quad (2)$$

$W$ is a binary up-sampling operator, indicating that zero is inserted between adjacent data points; $a_o$ and $b_i$ are wavelet low-pass and high-pass reconstruction filters, respectively, when $j = I, I-1, \cdots, 1$. The formula (2) of order $I$ is expressed by the operator $E^-$, that is, $v_E = E^- z_v$.

Step 5: Repeat steps 1, 3, and 4 repeatedly until the Baud sign extraction of all voxel time series data is complete.

## III. MIXING REMOVAL OF MULTI-SCALE FEATURE EXTRACTION BY DISCRETE WAVELET TRANSFORM

Mixing is a common problem in wavelet single-stage reconstruction, which is mainly manifested as follows: wavelet filtering effect is not good enough. Both drop sampling during decomposition and up sampling during the reconstruction will cause false frequencies [12-13]. For this purpose, the project intends to remove the drop sampling during decomposition and sample the top sampling during reconstruction. The use of a low-pass filter serves to eliminate high-frequency elements, while a high pass filter is applied to get rid of low-frequency components. This paper proposes an elimination method which is adapted to MATLAB wavelet, that is, keep the up-sampling and down-sampling, and then add a new filtering operation after the filtering of Mallat [14]. In more detail, the following formula (3) describes the algorithm for wavelet decomposition, but is still represented by $E$, i.e., $z_v = Ev$, with

$$\begin{cases} l_0 = v \\ l_j = R_0 G_K A_o l_{j-1} \\ r_j = R_0 G_B B_i l_{j-1} \end{cases} \quad (3)$$

$$\begin{cases} l_{j-1} = a_o G_K W l_j + b_i G_B W r_j \\ v_E = l_0 \end{cases} \quad (4)$$

$G_K$ and $G_B$ in equations (3) and (4) are re-filtering operators, $G_K v$ means that the coefficient of the middle half of the signal $v$ is set to zero after Fourier decomposition, and then Fourier reconstruction is carried out, and $G_B v$ means that the decomposition coefficient of the front and back 1/4 is set to zero during Fourier reconstruction [15]. The advantage of this is that $G_K$ and $G_B$ can be directly embedded in the wavelet transform function of MATLAB, thus taking full advantage of the advantages of MATLAB, such as data continuation in convolution [16]. The characteristic wavelet scale of $U$ is reconstructed by the multi-scale feature extraction method of discrete wavelet transform, and the spectrum of the reconstructed signal $U_E$ is obtained. Apparently mixing has been eradicated.

## IV. MULTI-SCALE FEATURE EXTRACTION OF DISCRETE WAVELET TRANSFORM BASED ON MATRIX ALGORITHM

This endeavor aims to devise a novel wavelet-based multiscale feature extraction technique, designed to address the limitations of conventional methods and streamline the computational burden associated with processing extensive functional MRI datasets[17]. Initially, the algorithm organizes the temporal data of each pixel in the FMRI sequentially. Subsequently, it structures this information into a matrix format for each pixel[18]. Thereafter, by leveraging matrix operations, the algorithm performs comprehensive feature extraction across the entire dataset in one go.

$$V_E = N_E V \quad (5)$$

Therefore, this project intends to transform the multi-scale feature extraction into M-wavelet transform multi-scale feature extraction, and transform the original wavelet multi-scale feature extraction into C wavelet transform to realize multi-scale feature extraction [19]. The utility of M-wavelet multiscale characteristics for EEG analysis is evaluated by employing signal activation identification techniques[20]. That is, the correlation coefficient $r$ between signal $U_E$ and each column vector in matrix $V_E$ is found, and Fisher's $C$ transformation is used

$$C = \frac{\sqrt{A-3}}{2} \ln \frac{1+r}{1-r} \quad (6)$$

For a training sample, an image classification process based on a DRBM sparse coding method is illustrated in Figure 2 (image cited in Systems 2023, 11(11), 547).

The first step: sampling the noise samples in the training sample set.

Step 2: The boundary of the training image is extracted by Canny operator, and the gray scale is added to the original image to obtain the image with texture strengthening effect.

Step 3: Read and sample from the texture-enhanced image to obtain a training feature base [21]. Let's call $R = \{r_1, r_2, \cdots, r_N\}$, $r$ a vector of 10,000 dimensions and $N$ the total number of features.

Step 4: The DRBM network is used to encode the features in the feature database, and CD, gradient decline and other methods are used to fuse with the class tags in the sample to obtain the optimal DRBM classifier[22].

Step 5: Input the image to be tested into the trained DRBM, and classify the image according to the obtained image to be identified[23].

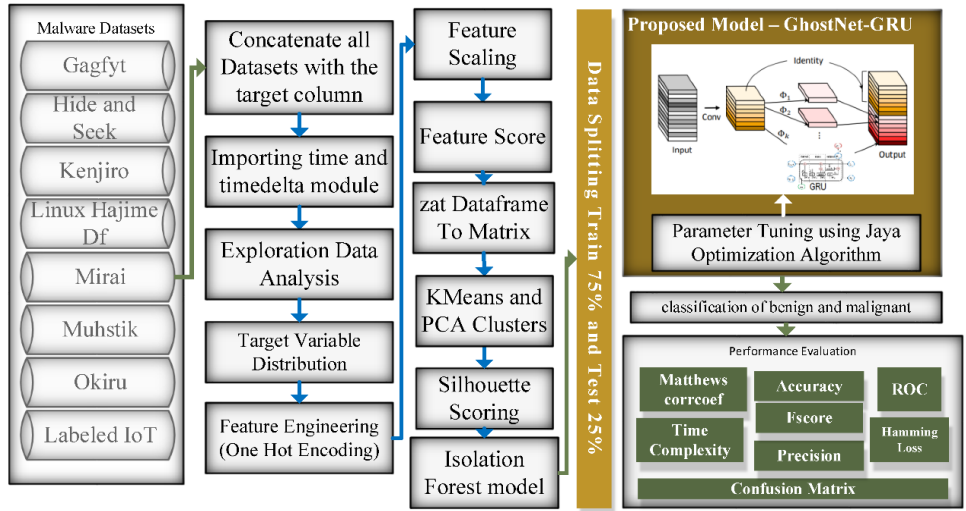

Fig. 2. Image classification process based on DRBM and edge detection

## V. SYSTEM SIMULATION

The data in this study are divided into two parts. The first one is collected from the medical forum and includes two types of brain images, both of which are 300x353. Additional information was collected from the web pages of the American Film Information and Data Center. Two sets of data made up the sample of the whole study, a total of 282 samples, of which 167 samples were normal and 115 samples were abnormal. Figure 3 is A partial MR Image of the brain, Figure 3A is a normal MR Image of the brain without lesions, and Figure 3(b) is an abnormal Image of the brain with lesions.

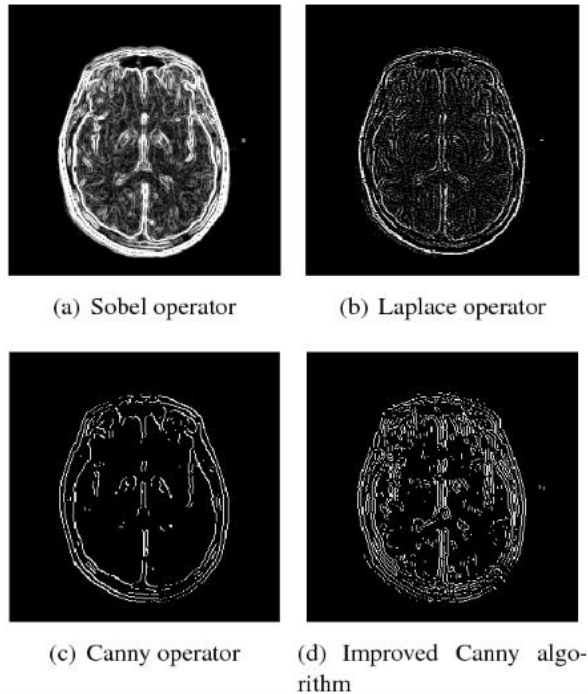

(a) Sobel operator  (b) Laplace operator
(c) Canny operator  (d) Improved Canny algorithm

Fig. 3. Magnetic resonance imaging of the brain

In view of the characteristics that MRI images are susceptible to noise during transmission, this study adopts an unsupervised learning approach which provided a new pathway for rapid and precise analysis of large-volume fMRI data[24] . This unsupervised learning strategy involves the use of a content encoder and a random noise encoder to segregate content information and noise artifacts within low-quality, noise-affected MRI images. We normalize noise distribution employing matrix calculations based on wavelet transformations and employ Kullback-Leibler divergence loss for regularization. This approach maintains the coherence of content information between the noisy inputs and the denoised outputs. Employing this methodology has led to notable enhancements in visual quality, offering a novel approach for the swift and accurate analysis of extensive fMRI datasets. Canny operator is used for boundary detection before boundary recognition and enhancement of image data[25]. The results obtained are shown in Figure 4. Figure 4(a) shows the unprocessed MR Image of the brain, 4(b) shows the boundary detection by the Canny operator, and 4(c) shows the enhanced image of the structure.

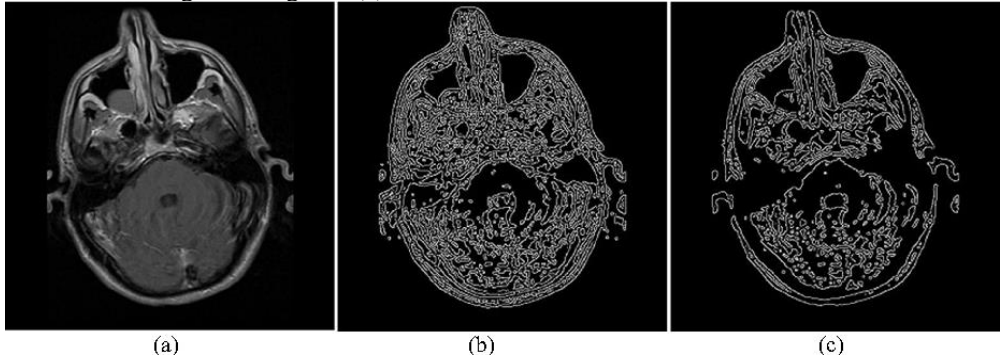

Fig. 4. Effect of edge detection by Canny operator

Experiments were conducted on 136 normally classified and 88 atypically classified brain MRI images, and other samples were examined. Each trial was conducted five times with a variety of randomly constructed training and test sets. The correct rate of each experiment is analyzed statistically, and the average and standard deviation of the correct rate of each experiment are obtained. In order to improve the learning speed, this paper adopts the method of down-sampling to sample the original image, so that the original image reaches 100x100. Among them, the input layer of DRBM is composed of 100x100+2 visual cells, and the number of hidden layers is set to 8500. When the neural network is initialized, it is initialized with a series of arbitrary values selected from the normal distribution, where the mean is 0 and the standard deviation is 1. Under different learning rates, the correct recognition results were obtained, as shown in Table 1.

TABLE I. COMPARISON OF CLASSIFICATION RESULTS UNDER LEARNING RATE

| Serial number | Learning rate | Accuracy rate |
|---|---|---|
| 1 | 0.104% | 93.45% |
| 2 | 0.313% | 91.92% |
| 3 | 0.521% | 88.84% |
| 4 | 0.729% | 87.31% |
| 5 | 1.042% | 84.25% |

The test results show that the sample is effectively identified in the learning rate range of 0.001~0.005, in which the learning rate of 0.001~0.005 is the best, which is 90.22%. The average accuracy is 87.25%. The proposed algorithm is compared with other commonly used image classification algorithms. The results obtained are shown in Table 2.

TABLE II. COMPARISON OF RESULTS OF DIFFERENT CLASSIFICATION METHODS

| Classification method | Accuracy rate |
|---|---|
| PCA+ rule-based association classifier | 81.19% |
| Support vector machine classifier | 85.78% |
| DRBM | 90.38% |
| Textual method | 93.45% |

The results show that compared with other algorithms, DRBM algorithm is superior to other algorithms in recognition results, which also proves that DRBM algorithm can better extract features from images, and the extracted features are more representative and selective. Through the boundary extraction and enhancement of the image, the noise is eliminated to the maximum extent, and the texture characteristics and recognition accuracy of the image are improved.

VI. CONCLUSION

DRBM is combined with boundary extraction to improve its recognition accuracy. Through boundary extraction and texture enhancement processing of multiple images, and fusion with DRBM neural network, better results can be obtained. This study has explored the synergy between the Deep Restricted Boltzmann Machine (DRBM) algorithm and image boundary extraction techniques to augment the precision of image recognition tasks. Our findings elucidate the DRBM algorithm's superior capability in feature extraction from images compared to other algorithms. Specifically, it demonstrates a heightened proficiency in deriving features that are both representative and selective, which are crucial for the enhancement of recognition results. The integration of boundary extraction and enhancement methodologies not only mitigates noise interference but also amplifies the distinctiveness of texture characteristics, thereby bolstering recognition accuracy. The implications of this research are multifaceted. For one, it opens avenues for further exploration into the optimization of feature extraction techniques and their integration with neural network models. Additionally, this study underscores the potential for developing more sophisticated image processing algorithms that can handle a broader spectrum of recognition scenarios, including those with highly cluttered backgrounds or low contrast.